\documentclass[prd,floats,floatfix,nofootinbib]{revtex4}
\usepackage[dvips]{graphicx}
\usepackage{graphics}
\usepackage{dcolumn}
\usepackage{amssymb}
\usepackage{bm}
\usepackage{amsmath}
\usepackage{color}
\usepackage{subfigure}
\usepackage{epstopdf}

\begin{document}

\title{General Boosted Black Holes:
A First Approximation}

\author{Rodrigo Maier\footnote{rodrigo.maier@uerj.br}} 

\affiliation{
Departamento de F\'isica Te\'orica, Instituto de F\'isica, Universidade do Estado do Rio de Janeiro,\\
Rua S\~ao Francisco Xavier 524, Maracan\~a,\\
CEP20550-900, Rio de Janeiro, Brazil
}

\date{\today}

\begin{abstract}
In this paper we obtain an approximate solution of Einstein field equations which describes a general boosted Kerr-Newman black hole relative to a Lorentz frame at future null infinity.
The boosted black hole is obtained from 
a general twisting metric whose boost emerges from the Bondi–Metzner–Sachs (BMS) group. Employing a standard procedure we build the electromagnetic energy-momentum
tensor with the Kerr boosted metric together with its timelike Killing vector as the electromagnetic potential. 
We demonstrate that our solution satisfies
Einstein field equations  up to a fourth-order expansion in 1/r, indicating that the spacetime closely resembles a Kerr-Newman black hole whose boost points in a arbitrary direction.
Spacetime structures of the general black hole -- namely the event horizon and ergosphere -- are examined in Bondi-Sachs coordinates.
For a proper timelike observer we show that the electric field 
generated by the boosted black hole
exhibits a purely radial behavior, whereas the magnetic field develops a complex structure characterized by two pronounced lobes oriented opposite to the boost direction.
\end{abstract}

\maketitle

\section{Introduction}
\label{intro}

The no-hair theorem remains a foundational principle in theoretical physics; however, empirical findings suggest that mass, angular momentum, and charge alone do not constitute a comprehensive set of parameters for describing the remnants of black holes formed through binary mergers. Recent analyses from the LIGO and Virgo collaborations\cite{virgo1,virgo2,virgo3,virgo4} reveal that these mergers occur with mass ratios ranging from 0.8 to 0.5, implying that the resulting black holes acquire a directional boost relative to an asymptotic Lorentz frame at null infinity, where gravitational wave signals are detected.

From a Kerr-Schild perspective\cite{Huq:2000qx,Madler:2017qlu}, boosted Schwarzschild black holes have been examined in the
literature. In \cite{Penna:2015qta} for instance, the author examines Blandford--Znajek\cite{Blandford:1977ds} power jets for boosted Schwarzschild black holes. 
It is shown that a power jet of a given boosted black hole is directly proportional to its velocity relative to the local compass of inertia. 
Gravitational radiation memory and its effect on symmetries of a body whose exterior is connected to a boosted Schwarzschild spacetime were also examined in \cite{Madler:2017umy}.

Extending the above scenario, the quest for boosted rotating black holes has been a subject of important debate. In this context, numerical relativity has played a significant role towards initial data of spinning binary black holes. In \cite{Karkowski:2006kp} for instance, initial data for boosted rotating black holes were constructed
in the axially symmetric case. 
Kerr-Schild metrics\cite{kramer} of the form $g_{\mu\nu}=\eta_{\mu\nu}+h k_\mu k_\nu$
have also played an important role considering its invariance under Lorentz symmetry about
the Minkowski background $\eta_{\mu\nu}$. Furthermore, still in the context of numerical relativity it has been shown that
boosted black holes from their Kerr-Schild versions may
furnish initial data for black holes in a binary orbit\cite{Huq:2000qx,Matzner:1998pt,Bonning:2003im,Cook:1997qc}.
Extending Wald's solution\cite{Wald:1974np}, in \cite{Morozova:2013ina} the authors obtain analytic solutions of Maxwell equations for a boosted rotating black hole embedded in an asymptotically uniform magnetic test field. In this context, energy losses due to particles accelerated along the magnetic field lines were evaluated. For the last, but not least, in \cite{Balasin:1995tj,Barrabes:2003up} the authors break axial symmetry by applying a Lorentz boost on Kerr-Schild coordinates of a Kerr black hole. 

The methodologies outlined above are rather different from those explored in references \cite{Soares:2016bte,Soares:2018yym,Gallo:2019emy,Soares:2020gqy,Aranha:2021zwf}. 
In this context, a general twisting metric whose boost emerges from 
the Bondi–Metzner–Sachs (BMS) group within the characteristic initial value framework is obtained.
By the time when \cite{Soares:2016bte,Soares:2018yym} were published, it was initially believed that exact Kerr metrics whose boost is given by the BMS group had been established. However, in \cite{Gallo:2019emy} the authors identified inconsistencies in the boosted solutions of 
\cite{Soares:2016bte,Soares:2018yym}, showing that a transformation of angular coordinates applied to the original Kerr metric could produce these alleged boosted configurations, thereby challenging their validity. These issues were subsequently addressed in \cite{Soares:2020gqy}, where the author derived the complete asymptotic Lorentz transformations between Robinson–Trautman and Bondi–Sachs coordinates, accounting for the perturbative contribution of the Lense–Thirring effect arising from Kerr rotation. This analytical development resulted in an approximate model of a boosted, axially symmetric Kerr black hole, assessed from a Lorentz frame at future null infinity. A broader generalization -- where the boost is not necessarily aligned with the rotational axis -- is presented in \cite{Aranha:2022fuq}, describing a spacetime expected to typify the remnant black hole formed through the merger of uncharged, rotating black holes. These approximate solutions incorporate three independent parameters and define a transformation within the generalized BMS group, as formulated in \cite{Bondi:1962px} and examined in detail by Sachs\cite{Sachs:1962zza}. This transformation exemplifies the general framework of Lorentz boosts embedded in the homogeneous Lorentz sector of the BMS group at future null infinity.

In addition to the advancements mentioned above, the assumption of a nonzero electric charge introduces a potentially pivotal element in modeling more general black hole configurations. While electrically charged black holes are generally thought to be uncommon
-- primarily due to their propensity to attract and neutralize opposite charges from their neighborhoods -- substantial efforts have been devoted to analyze processes that regulate this neutralization (see \cite{Eardley:1975kp,Ruffini:2009hg,Hwang:2010im,Gong:2019aqa} and cited references). From an astrophysical perspective, it has been hypothesized that binary black hole systems may retain residual charge. According to Zhang’s mechanism\cite{Zhang:2016rli}, a rotating charged black hole can form a magnetosphere, allowing it to sustain an electric charge over extended periods. This phenomenon may produce electromagnetic emissions during gravitational wave events triggered by binary black hole mergers. Additionally, the investigation in \cite{Zajacek:2018ycb} examines the charge of the Galactic supermassive black hole, offering plausible bounds on its magnitude and the relevant temporal scales for charge acquisition and dissipation. Further studies \cite{Liebling:2016orx,Liu:2016olx,Punsly:2016abn,Levin:2018mzg,Deng:2018wmy} lend continued support to the existence of charged black holes, particularly in connection with electromagnetic signals observed in merger events.
Following a route analogous to that of shown in \cite{Soares:2020gqy,Aranha:2021zwf}, in the reference \cite{Aranha:2024aye} the authors obtain an approximate Kerr-Newman solution whose boost -- 
which is aligned with the black hole angular momentum --
is given by the BMS group within the characteristic initial value framework.
In this paper we extend the analysis presented in \cite{Aranha:2024aye} 
in order to obtain the approximate metric of a general charged black hole whose boost points in a arbitrary direction relative to its rotation axis.

In the next section we obtain our general approximate solution.
To this end we follow the
standard procedure used in \cite{Aranha:2024aye,Maier:2023xhl} which shows that
the energy-momentum tensor of
a Kerr-Newman black hole can be built with
the Kerr metric together with its timelike Killing vector as the electromagnetic
potential -- the Papapetrou field\cite{Papapetrou:1966zz}. 
In section $3$ we examine spacetime structures -- namely the event horizon and ergosphere -- which organize the causal structure of our general solution. 
In section $4$ the electromagnetic pattern of our approximate solution is examined.
We leave our final remarks in section $5$.

\section{General Boosted Black Holes}

\subsection{Kerr Boosted Black Holes}

We start by considering the Kerr metric
\cite{Kerr:1963ud}
in its traditional Eddington-Finkelstein form with coordinates $(u, r, \chi. \phi)$:
\begin{eqnarray}
\label{eqko}
ds^2=(r^2+\omega_0^2 \cos^2\chi)(d\chi^2+\sin^2\chi d\phi^2)
&-&2(du+\omega_0\sin^2\chi d\phi)(dr-\omega_0\sin^2\chi d\phi)\\
\nonumber
&&-\Big(1-\frac{2 m_0 r}{r^2+\omega_0^2\cos^2\chi}\Big)(du+\omega_0\sin^2\chi d\phi)^2.
\end{eqnarray}
In the above the parameters $m_0$ and $\omega_0$ stand for the black hole mass and angular momentum. respectively.
Applying the angular transformation
\begin{eqnarray}
\label{tg}
\cos\chi=\frac{b+a\cos\theta}{\kappa(\theta)}    
\end{eqnarray}
where 
\begin{eqnarray}
\kappa(\theta)=a+b\cos\theta,    
\end{eqnarray}
and 
$a=\cosh\gamma$ and $b=\sinh\gamma$, we end up with 
\begin{eqnarray}
\label{eqi0}
ds^2=\frac{r^2+\sigma^2(\theta)}{\kappa^2(\theta)}(d\theta^2+\sin^2\theta d\phi^2)
&-&2\Big[du+\frac{\omega_0\sin^2\theta}{\kappa^2(\theta)}d\phi\Big]\Big[dr-\frac{\omega_0\sin^2\theta}{\kappa^2(\theta)}d\phi\Big]\\
\nonumber
&&-\Big(1-\frac{2m_0 r}{r^2+\sigma^2(\theta)}\Big)\Big(du+\frac{\omega_0\sin^2\theta}{\kappa^2(\theta)}d\phi\Big)^2,
\end{eqnarray}
where
\begin{eqnarray}
\sigma(\theta)=\omega_0\Big[\frac{b+a\cos\theta}{\kappa(\theta)}\Big].    
\end{eqnarray}
The metric (\ref{eqi0}) was obtained in \cite{Soares:2016bte} under the assumption that $\gamma$ represents the boost intensity parameter of a boosted Kerr black hole. However, in \cite{Gallo:2019emy} the authors noted 
-- in the same manner shown above --
that the specific angular transformation (\ref{tg}), when applied to the original Kerr metric (\ref{eqko}), reproduces the alleged boosted configurations (\ref{eqi0}) obtained in \cite{Soares:2016bte}. This observation casts doubt on the authenticity of (\ref{eqi0}) as a genuinely boosted solution. 
The inconsistencies identified in \cite{Gallo:2019emy} were subsequently examined in \cite{Soares:2020gqy}, wherein the author derived the full set of asymptotic Lorentz transformations connecting Robinson–Trautman and Bondi–Sachs coordinate systems. This derivation incorporated the perturbative effects of the Lense–Thirring precession induced by the rotational dynamics of the Kerr metric. As a result, an approximate representation of a boosted, axially symmetric Kerr black hole was formulated, as viewed from a Lorentz frame positioned at future null infinity. 

For stationary axisymmetric configurations,
the transformations from Robinson–Trautman coordinates $(u, r, \theta, \phi)$ to Bondi–Sachs coordinates $(U, R, \theta, \phi)$ -- under conditions satisfying the appropriate Bondi–Sachs asymptotic framework -- are given by\cite{Gallo:2019emy,Soares:2020gqy,Aranha:2013rj} 
\begin{eqnarray}
\label{rit1n0}
r=\kappa(\theta) R,~~ dr\simeq \kappa(\theta) dR, ~~du\simeq\frac{dU}{\kappa(\theta)},
\end{eqnarray}
together with a Bondi mass aspect expressed as
\begin{eqnarray}
m_b \simeq \frac{m_0}{\kappa^3(\theta)}.    
\end{eqnarray}
Applying (\ref{rit1n0}) to (\ref{eqi0}) we obtain the original form the line element obtained in \cite{Soares:2020gqy}, namely
\begin{eqnarray}
\label{bsiv}
ds^2=[R^2+\sigma^2(\theta)](d\theta^2+\sin^2\theta d\phi^2)
&-&2(dU+\omega_b\sin^2\theta d\phi)\Big[dR-\frac{\omega_b\sin^2\theta}{\kappa^2(\theta)}d\phi\Big]\\
\nonumber
&&-(dU+\omega_b\sin^2\theta d\phi)^2\Big[\frac{1}{\kappa^2(\theta)}-\frac{2m_b R}{R^2+\sigma^2(\theta)}\Big]+\mathcal{O}\Big(\frac{1}{R^2}\Big),
\end{eqnarray}
where the rotation parameter transform as
\begin{eqnarray}
&\omega_0&\rightarrow \omega_b=\frac{\omega_0}{\kappa(\theta)}.   
\end{eqnarray}
In this context, (\ref{bsiv}) was understood as an approximate model of a boosted, axially symmetric Kerr black hole perceived from a Lorentz frame situated at future null infinity.

The analysis performed in \cite{Soares:2016bte} was extended in \cite{Soares:2018yym} 
using the same framework presented in \cite{kramer}.
In this case, a more general configuration in which the supposed boost is not aligned with the black hole angular momentum
was obtained so that (\ref{eqi0}) turns into
\begin{eqnarray}
\nonumber
ds^2&=&\frac{r^2+\Sigma^2(\theta, \phi)}{K^2(\theta, \phi)}(d\theta^2+\sin^2{\theta}d\phi^2)
-2[du-2{\cal L}(\theta, \phi)\cot{(\theta/2)}d\phi]\\
\label{e1}
\nonumber
&&\times \Big\{dr+ \frac{\omega_0}{K^2(\theta, \phi)}\Big[({n_2\sin{\phi}-n_3\cos{\phi}})d\theta
\\
&&~~~~~~~~~~~~+\Big({(n_2\cos{\phi}+n_3\sin{\phi})\sin\theta\cos\theta-n_1 \sin^2{\theta}}\Big)d\phi\Big] \Big\} \\
\nonumber
&&~~~~~~~~~~~~~~~~~~-[du-2{\cal L}(\theta, \phi)\cot{(\theta/2)}d\phi]^2
\Big[1-\frac{2m_0 r}{r^2+\Sigma^2(\theta, \phi)}\Big].
\end{eqnarray}
In the above the function ${\cal L}(\theta, \phi)$ satisfies the differential equation
\begin{eqnarray}
\label{eqL}
{\cal L}_{,\theta}-\frac{{\cal L}}{\sin\theta}+(1-\cos\theta)\frac{\Sigma(\theta, \phi)}{K^2(\theta, \phi)}=0
\end{eqnarray}
and
\begin{eqnarray}
\label{e2}
K(\theta, \phi)&=&a+b(\boldsymbol{\hat{x}}\cdot\boldsymbol {\hat{n}}),\\
\Sigma(\theta, \phi)&=&\frac{\omega_0}{K(\theta, \phi)}[b+a({\boldsymbol{\hat{x}}}\cdot\boldsymbol {\hat{n}})].
\end{eqnarray}
Finally, ${\boldsymbol {\hat{x}}}\equiv (\cos\theta, \sin\theta\cos \phi,\sin\theta\sin\phi)$ is the unit vector
along an arbitrary direction and $\boldsymbol{\hat{n}} = (n_1, n_2, n_3)$ 
is the boost direction satisfying
\begin{eqnarray}
n_1^2+n_2^2+n_3^2=1.
\end{eqnarray}

However, it is easy to see that for $n_2=n_3=0$, (\ref{e1}) reduces to (\ref{eqi0}) and, as discussed above, 
the original Kerr metric in its Eddington-Finkelstein form is recovered. In this sense -- as also noted by \cite{Gallo:2019emy} --
the exact solution (\ref{e1}) should not provide a proper boosted Kerr metric. In order to circumvent such issue we now generalize the standard procedure presented in \cite{Soares:2020gqy} by extending the transformations (\ref{rit1n0}) as   
\begin{eqnarray}
\label{rit1n}
r=K(\theta, \phi) R,~~ dr\simeq K(\theta, \phi) dR, ~~du\simeq\frac{dU}{K(\theta, \phi)}.
\end{eqnarray}
After applying (\ref{rit1n}) to the exact metric (\ref{e1}) 
%together with the substitution of the Bondi aspect mass
%given by
%
%\begin{eqnarray}
%m_b\simeq \frac{m_0}{K^3(\theta, \phi)},    
%\end{eqnarray}
%
it can then be shown\cite{Aranha:2022fuq} that we end up with a boosted metric which asymptotically assumes the following form in the Bondi-Sachs coordinates $(U, R, \theta, \phi)$:
\begin{eqnarray}
\label{bsbhk}
\nonumber
ds^2=\Big[R^2+\frac{\Sigma^2(\theta, \phi)}{K^2(\theta, \phi)}\Big](d\theta^2+\sin^2\theta d\phi^2)-2\Big[\frac{dU}{K(\theta, \phi)}
-2{\cal L}(\theta, \phi)\cot(\theta/2)d\phi\Big]\\
\times\Big\{K(\theta, \phi)dR
+ \omega_0\Big[\frac{n_2\sin{\phi}-n_3\cos{\phi}}{K^2(\theta, \phi)}\Big]d\theta+\omega_0\Big[\frac{(n_2\cos{\phi}+n_3\sin{\phi})\sin\theta\cos\theta-n_1 \sin^2{\theta}}{K^2(\theta, \phi)}\Big]d\phi \Big\}
\\
\nonumber
-\Big[\frac{dU}{K(\theta, \phi)}-2{\cal L}(\theta, \phi)\cot(\theta/2)d\phi\Big]^2
\Big[1-\frac{2m_0 R K(\theta,\phi)}{R^2K^2(\theta,\phi)+\Sigma^2(\theta, \phi)}   \Big]+\mathcal{O}\Big(\frac{1}{R^2}\Big).
\end{eqnarray}
At this stage is worth to mention that
\begin{eqnarray}
\label{eqgrt}
g^{R\theta}=\mathcal{O}\Big(\frac{1}{R^2}\Big)=g^{R\phi},
\end{eqnarray}
thus satisfying Bondi-Sachs boundary conditions\cite{Aranha:2013rj} as one should expect.
The metric (\ref{bsbhk}) is an extension of the one obtained in \cite{Soares:2020gqy} in the sense that it describes general Kerr black hole whose boost is not necessarily align with its rotation axis. 

It is important to emphasize that once the metric (\ref{bsbhk}) is seen from a Lorentz frame located at future null infinity, our approximation scheme mediated by the transformations (\ref{rit1n}) becomes gauge-dependent and tailored for the Kerr metric -- the same remark applies to the axially symmetric metric (\ref{bsiv}) and transformations (\ref{rit1n0}). Notably, the Bondi–Sachs boundary conditions, which enforce asymptotic flatness and the presence of outgoing radiation, are designed to provide a comprehensive framework for integrating the field equations. In light of its formulation for radiative systems (see \cite{Aranha:2013rj} and references therein), one can verify that equation (\ref{eqgrt}) satisfies these boundary conditions for our particular configuration.

\subsection{ADM Decomposition}

In order to obtain the timelike Killing vector field of (\ref{bsbhk})
we now choose a preferred timelike foliation following the $1+3$ ADM decomposition. To this end 
we use (\ref{rit1n}) together with $du=dt-dr$ in (\ref{bsbhk}). In the coordinates $(t, r, \theta, \phi)$ the metric (\ref{bsbhk}) then turns into
\begin{eqnarray}
\label{adm}
ds^2=-N^2 dt^2+\gamma_{ij}(dx^i-N^idt)(dx^j-N^jdt)+\mathcal{O}\Big(\frac{1}{r^2}\Big),
\end{eqnarray}
where the components of the spatial metric are given by
\begin{eqnarray}
\gamma_{rr}&=&1+\frac{2m_0 r}{r^2+\Sigma^2(\theta, \phi)},\\
\gamma_{r\theta}&=&\frac{\omega_0}{K^2(\theta, \phi)}(n_2\sin\phi-n_3\cos\phi),\\
\gamma_{r\phi}&=&\frac{4m_0r\cot(\theta/2){\cal L(\theta, \phi)}}{r^2+\Sigma^2(\theta, \phi)}+\frac{\omega_0\sin\theta}{K^2(\theta, \phi)}[\cos\theta(n_2\cos\phi+n_3\sin\phi)-n_1\sin\theta],\\
\gamma_{\theta\theta}&=&\frac{r^2+\Sigma^2(\theta, \phi)}{K^2(\theta, \phi)},\\
\gamma_{\theta \phi}&=&\frac{2\omega_0\cot(\theta/2){\cal L(\theta, \phi)}}{K^2(\theta, \phi)}(n_2\sin\phi-n_3\cos\phi),
\end{eqnarray}
\begin{eqnarray}
\nonumber
\gamma_{\phi\phi}&=&-4\cot^2(\theta/2){\cal L}^2(\theta,\phi)\Big[1-\frac{2m_0r}{r^2+\Sigma^2(\theta,\phi)}\Big]\\
&+&\frac{2\cos^2(\theta/2)}{K^2(\theta,\phi)}\Big\{(1-\cos\theta)[r^2+\Sigma^2(\theta,\phi)]
+4\omega_0{\cal L}(\theta,\phi)(n_2\cos\theta\cos\phi-n_1\sin\theta+n_3\cos\theta\sin\phi)\Big\}.
\end{eqnarray}
The shift components on the other hand read,
\begin{eqnarray}
N^r &=&1-\frac{\gamma_{\theta\theta}\gamma_{\phi\phi}-\gamma^2_{\theta\phi}-2(\gamma_{r\phi}\gamma_{\theta\theta}-
\gamma_{r\theta}\gamma_{\theta\phi})\cot(\theta/2){\cal L}}{\tilde{\gamma}},\\
N^\theta &=&\frac{\gamma_{r\theta}\gamma_{\phi\phi}-\gamma_{r\phi}\gamma_{\theta\phi}+2(\gamma_{rr}\gamma_{\theta\phi}-\gamma_{r\theta}\gamma_{r\phi})\cot(\theta/2){\cal L}}{\tilde{\gamma}},\\
N^\phi &=&-\frac{\gamma_{r\theta}\gamma_{\theta\phi}-\gamma_{r\phi}\gamma_{\theta\theta}-2(\gamma^2_{r\theta}-\gamma_{rr}\gamma_{\theta\theta})\cot(\theta/2){\cal L}}{\tilde{\gamma}},
\end{eqnarray}
where $\tilde{\gamma}\equiv {\rm det}(\gamma_{ij})$. Finally, the lapse function is given by
\begin{eqnarray}
\label{lapse}
N=\pm\sqrt{1+\gamma_{ij}N^iN^j-\frac{2rm_0}{r^2+\Sigma^2(\theta, \phi)}}.
\end{eqnarray}
It is then easy to see that $T^\mu\equiv \epsilon_0\delta^\mu_{~t}$ 
-- where $\epsilon_0$ is a constant parameter --
satisfies the Killing equation so that $T^\mu$ may be understood as 
an electromagnetic potential which satisfies the Maxwell equations\cite{Wald:1974np,Papapetrou:1966zz} 
in the background described by (\ref{adm})--(\ref{lapse}).

\subsection{Boosted Kerr-Newman Black Holes}

To proceed we now apply our standard procedure to build the electromagnetic energy-momentum tensor with the Kerr boosted metric together with its timelike Killing vector as the electromagnetic potential. To this end we start noting that the timelike Killing vector of (\ref{adm}) reads $T^\mu\equiv \epsilon_0\delta^\mu_{~t}$, where $\epsilon_0$ is an arbitrary constant. Once (\ref{adm}) describes an empty spacetime, it is easy to show\cite{Wald:1974np,Papapetrou:1966zz} that
the Faraday tensor built with $T^\mu$, that is, 
\begin{eqnarray}
\label{faraday}
F_{\mu\nu}=\nabla_\mu T_\nu-\nabla_\nu T_\mu\equiv-2\nabla_\nu T_\mu    
\end{eqnarray}
furnishes a solution of Maxwell equations in the
background metric (\ref{adm}).

Following the prescription (\ref{faraday}) we construct the electromagnetic energy-momentum tensor
\begin{eqnarray}
T_{\mu\nu}=F_{\mu}^{~\alpha}F_{\nu\alpha}-\frac{1}{4}g_{\mu\nu}F^{\alpha\beta}F_{\alpha\beta} 
\end{eqnarray}
with the background metric (\ref{adm}). It can then be shown that the nonvanishing components of $T^{\mu}_{~\nu}$ (up to 4th order in $1/r$) read
\begin{eqnarray}
\label{tm}
&&T^t_{~t}\simeq -\frac{2m_0^2\epsilon_0^2}{r^4},~~T^t_{~\phi}\simeq \frac{8m_0^2\epsilon_0^2}{r^4}\cot(\theta/2){\cal L}(\theta, \phi),\\
\label{tm1}
&&T^r_{~r}\simeq -\frac{2m_0^2\epsilon_0^2}{r^4},~~T^\theta_{~\theta}\simeq T^\phi_{~\phi}\simeq\frac{2m_0^2\epsilon_0^2}{r^4}.
\end{eqnarray}

In order to obtain the general Kerr-Newman metric of a
boosted black hole we now consider the natural extension of (\ref{e1}) employing the substitution
$2m_0 r \rightarrow  2m_0 r - q_0^2$, that is
\begin{eqnarray}
\nonumber
ds^2&=&\frac{r^2+\Sigma^2(\theta, \phi)}{K^2(\theta, \phi)}(d\theta^2+\sin^2{\theta}d\phi^2)
-2[du-2{\cal L}(\theta, \phi)\cot{(\theta/2)}d\phi]\\
\label{e1n}
\nonumber
&&\times \Big\{dr+ \frac{\omega_0}{K^2(\theta, \phi)}\Big[({n_2\sin{\phi}-n_3\cos{\phi}})d\theta
\\
&&~~~~~~~~~~~~+\Big({(n_2\cos{\phi}+n_3\sin{\phi})\sin\theta\cos\theta-n_1 \sin^2{\theta}}\Big)d\phi\Big] \Big\} \\
\nonumber
&&~~~~~~~~~~~~~~~~~~-[du-2{\cal L}(\theta, \phi)\cot{(\theta/2)}d\phi]^2
\Big[1-\frac{2m_0 r-q_0^2}{r^2+\Sigma^2(\theta, \phi)}\Big].
\end{eqnarray}
Applying the transformations (\ref{rit1n}) together with $du=dt-dr$,
we then obtain an approximate
boosted Kerr-Newman metric which, in the coordinates $(t, r, \theta, \phi)$, reads
\begin{eqnarray}
\nonumber
ds^2=-\Big[1-\frac{2m_0 r-q_0^2}{r^2+\Sigma^2(\theta,\phi)}\Big]dt^2-2\Big[\frac{2m_0 r-q_0^2}{r^2+\Sigma^2(\theta,\phi)}\Big]dtdr
+\frac{2\omega_0(n_3\cos\phi-n_2\sin\phi)}{K^2(\theta,\phi)}dtd\theta\\
\nonumber
+2\Big\{\frac{\omega_0\sin\theta[n_1\sin\theta-\cos\theta(n_2\cos\phi+n_3\sin\phi)]}{K^2(\theta,\phi)}
+2\cot(\theta/2){\cal L}(\theta,\phi)\Big[1-\frac{2m_0 r-q_0^2}{r^2+\Sigma^2(\theta,\phi)}\Big]\Big\}dtd\phi\\
\nonumber
+\Big[1+\frac{2m_0 r-q_0^2}{r^2+\Sigma^2(\theta,\phi)}\Big]dr^2-\frac{2\omega_0(n_3\cos\phi-n_2\sin\phi)}{K^2(\theta,\phi)}drd\theta\\
\nonumber
-2\Big\{\frac{\omega_0\sin\theta[n_1\sin\theta-\cos\theta(n_2\cos\phi+n_3\sin\phi)]}{K^2(\theta,\phi)}-2\cot(\theta/2){\cal L}(\theta,\phi)\Big[\frac{2m_0 r-q_0^2}{r^2+\Sigma^2(\theta,\phi)}\Big]\Big\}drd\phi\\
\nonumber
+\frac{r^2+\Sigma^2(\theta,\phi)}{K^2(\theta,\phi)}d\theta^2-\frac{4\omega_0\cot(\theta/2){\cal L}(\theta,\phi)}{K^2(\theta, \phi)}(n_3\cos\phi-n_2\sin\phi)d\theta d\phi\\
\label{gbbh}
-2\cot^2(\theta/2)\Big\{2{\cal L}^2(\theta,\phi)\Big[1-\frac{2m_0 r-q_0^2}{r^2+\Sigma^2(\theta,\phi)}\Big]
-\frac{1}{K^2(\theta,\phi)}[(1-\cos\theta)(r^2+\Sigma^2(\theta,\phi))~~\\
\nonumber
+4\omega_0{\cal L}(\theta,\phi)(n_2\cos\theta\cos\phi-n_1\sin\theta+n_3\cos\theta\sin\phi)]\Big\}d\phi^2+\mathcal{O}\Big(\frac{1}{r^2}\Big).
\end{eqnarray}

It can then be shown that the nonvanishing components
-- up to 4th order in 1/r -- of the Einstein tensor of (\ref{gbbh}) read
\begin{eqnarray}
\label{ein}
G^t_{~t}&\simeq& -\frac{q_0^2}{r^4},~~G^t_{~\phi}\simeq \frac{4 q^2_0}{r^4}\cot(\theta/2){\cal L}(\theta, \phi),\\
\label{ein1}
G^r_{~r}&\simeq& -\frac{q_0^2}{r^4},~~G^\theta_{~\theta}\simeq G^\phi_{~\phi}\simeq\frac{q_0^2}{r^4}.
\end{eqnarray}
Therefore, it is easy to see from (\ref{tm})--(\ref{tm1}) and (\ref{ein})--(\ref{ein1}) that Einstein field equations
\begin{eqnarray}
G^\mu_{~\nu}=\kappa^2T^{\mu}_{~\nu}    
\end{eqnarray}
are satisfied -- up to 4th order in 1/r -- once one fixes
\begin{eqnarray}
\label{ep}
\epsilon_0=\frac{q_0}{\sqrt{2}\kappa m_0}.
\end{eqnarray}

\section{Spacetime Structures}

We now examine spacetime structures of the general boosted black hole,
namely, its event horizon and ergosphere. To this end it is appropriate to write
(\ref{gbbh}) in Bondi-Sachs coordinates $(U, R, \theta, \phi)$ -- with the use of (\ref{rit1n}) -- so that
the coordinate singularities are given by the condition $g^{RR} = 0$. In this case it can be shown that the event horizon is located at
\begin{figure}
\begin{center}
\includegraphics[width=8cm,height=8.5cm]{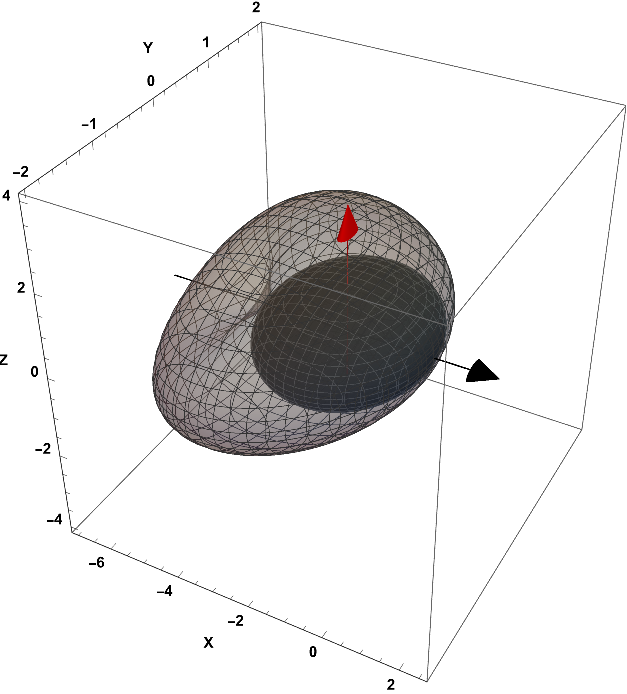}\includegraphics[width=8cm,height=8.5cm]{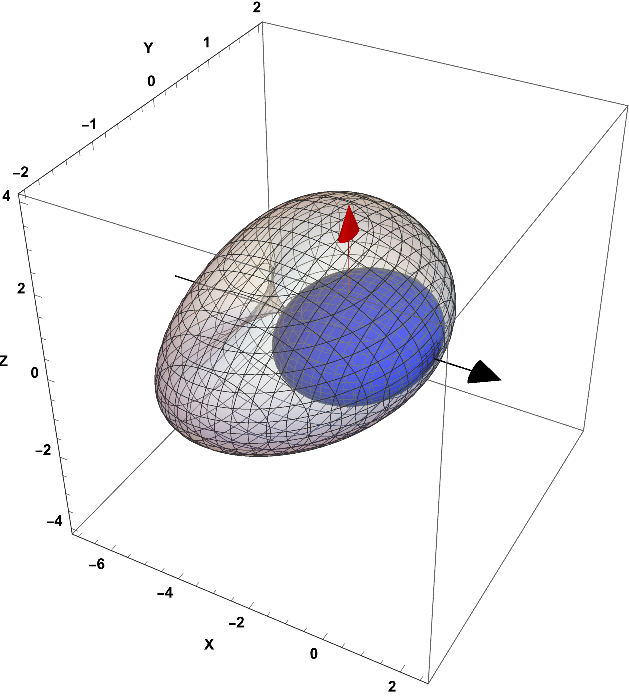}
\caption{The event horizons and ergospheres of boosted black holes -- solid and shaded surfaces, respectively -- in Bondi-Sachs coordinates. Thin red arrows indicate
the rotation axes while the black ones 
point towards boost directions.
In both plots we have fixed the parameters
$m_0=1.00$, $\omega_0=0.99$, $\gamma=1.30$, $n_1=n_3=0$ and $n_2=1$ so that the boost lies in the $x$ axis.
In the left panel we assume a vanishing electric charge and a boosted Kerr black hole is shown. In the right panel we have fixed 
$q_0=0.14106$. Here we explicitly see that a nonvanishing electric
charge diminishes the overall domain of the event horizon (hence also the ergosphere) when compared to neutral configurations.
Moreover, in both panels
we see an explicit deformation of the event horizons and ergospheres in the
boost direction. In this example it may be noticed that the ergospheres described by (\ref{ergo}) are indeed 
axially symmetric with respect
to the boost axis.
}
\label{fig1}
\end{center}
\end{figure}
\begin{eqnarray}
\label{hz}
R_h=\frac{m_0}{K(\theta,\phi)}+\frac{\sqrt{K^2(\theta,\phi)[m_0^2-q_0^2-\Sigma^2(\theta, \phi)]-\omega_0^2 Z^2(\theta, \phi)}}{K^2(\theta,\phi)}    
\end{eqnarray}
where
\begin{eqnarray}
Z^2(\theta, \phi)\equiv \sin^2\theta(n_1-n_2\cot\theta\cos\phi-n_3\cot\theta\sin\phi)^2+(n_3\cos\phi-n_2\sin\phi)^2.
\end{eqnarray}
From the above one notices that the event horizon formation condition is given by
\begin{eqnarray}
K^2(\theta,\phi)[m_0^2-q_0^2-\Sigma^2(\theta, \phi)]-\omega_0^2 Z(\theta, \phi)\geq 0.
\end{eqnarray}
The ergosphere on the other hand is defined by $g_{UU} = 0$. In Bondi-Sachs coordinates
we obtain
\begin{eqnarray}
\label{ergo}
R_{e}=\frac{m_0+\sqrt{m_0^2-q_0^2-\Sigma^2(\theta, \phi)}}{K(\theta, \phi)}.    
\end{eqnarray}
From the above we see that the horizon formation is a sufficient condition for the existence of an outer ergosphere, as one should expect. As discussed in \cite{Aranha:2022fuq}, it can be shown that the surfaces described by (\ref{ergo}) are 
axially symmetric with respect
to the boost axis. Moreover, it is easy to see that the ultimate effect of a nonvanishing electric charge is to shrink the
overall size of the event horizon and ergosphere areas when compared to neutral boosted Kerr black holes\cite{Aranha:2024aye}. 
In fact, fixing the black hole mass, angular momentum and boost parameters, from (\ref{hz}) and (\ref{ergo}) one may notice that $R_h$ and $R_e$ decrease as $q_0$ increases.
Such features are illustrated in Fig. 1. 

Although these numerical results correspond to mere
mathematical illustrations, it can be shown that the features mentioned above hold for the whole domain of the parametric space as long as an event horizon is formed. 

\section{Basic Electrodynamics}

To proceed we now examine the electromagnetic field measured from a proper timelike observer in the coordinates $(t, r, \theta, \phi)$ of (\ref{gbbh}). To this end we fix our observer four velocity as 
\begin{eqnarray}
\label{obs}
u^\mu=\frac{1}{\sqrt{-g_{tt}}}\delta^\mu_{~t}    
\end{eqnarray}
so that electric and magnetic
fields are given by using their usual definitions
\begin{eqnarray}
\label{emf}
E^\mu=F^{\mu}_{~\nu}u^\nu,\\
\label{emf2}
B^\mu={\cal F}^{\mu}_{~\nu}u^\nu.    
\end{eqnarray}
In the above $F^{\mu}_{~\nu}$ stands for the usual Faraday tensor while
${\cal F}^{\mu}_{~\nu}$ is its dual form, namely, 
\begin{eqnarray}
{\cal F}^{\mu\nu}\equiv \frac{1}{2}\epsilon^{\mu\nu\alpha\beta}F_{\alpha\beta}.    
\end{eqnarray}

\begin{figure}
\begin{center}
\includegraphics[width=7cm,height=6cm]{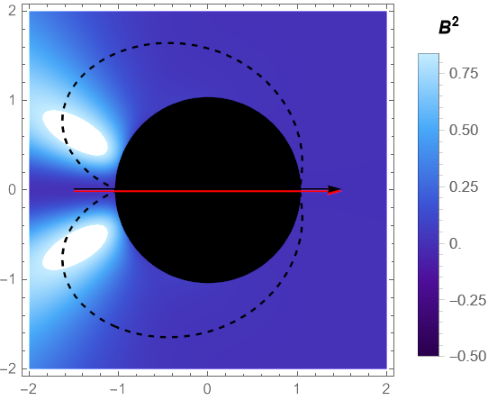}
\includegraphics[width=7cm,height=6cm]{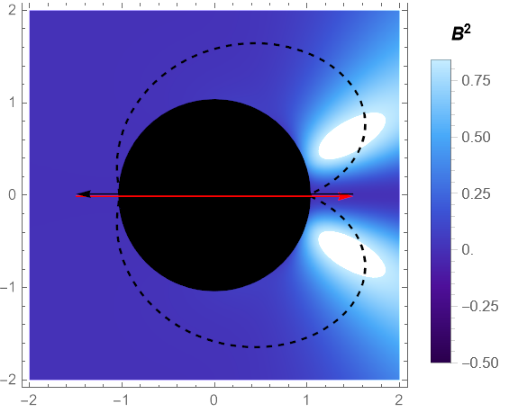}\\
\includegraphics[width=7cm,height=6cm]{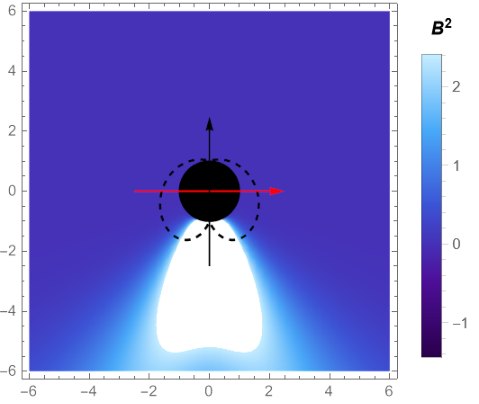}
\includegraphics[width=7cm,height=6cm]{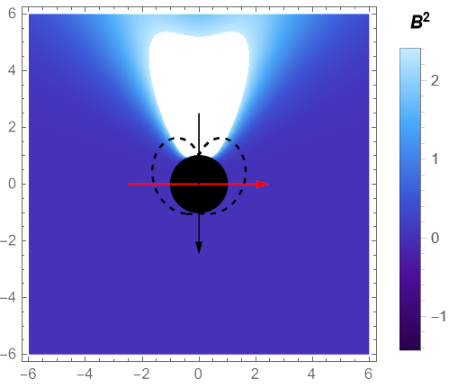}
\caption{The magnitude of the magnetic field for four different boost directions: ${\bf{\hat{n}}}=(1,0,0)$, top left; ${\bf{\hat{n}}}=(0,1,0)$, bottom left; ${\bf{\hat{n}}}=(-1,0,0)$, top right; ${\bf{\hat{n}}}=(0,-1,0)$, bottom right. In all these plots we have chosen the section $\phi=0$. Black arrows stand for the boost directions while the red ones indicate the black hole angular momentum. The solid black disks correspond to the inner region encapsulated by the event horizon and the dashed curves
are connected to the black hole ergosphere.
In these figures we have fixed the parameters $m_0=1.0$, $\omega_0=0.9$, $\gamma=1.0$, $q_0=0.435$ and $\kappa=1$.
}
\label{fig1}
\end{center}
\end{figure}

From (\ref{faraday})  it can be shown that the nonvanishing components of $F_{\mu\nu}$ -- up to second order in $1/r$ -- are given by
\begin{eqnarray}
\label{fdd1}
F_{tr}&\simeq&\frac{\sqrt{2}q_0}{\kappa r^2},\\
\label{fdd2}
F_{r\phi}&\simeq&\frac{2\sqrt{2}q_0}{\kappa r^2}\cot{(\theta/2)}{\cal L}(\theta, \phi),\\
\nonumber
F_{\theta\phi}&\simeq& \frac{\sqrt{2}q_0}{\kappa m_0}\Big\{  
\frac{{\cal L}(\theta, \phi)}{\cos\theta-1}+\cot(\theta/2){\cal L},_{\theta}
+\frac{\omega_0}{K^3(\theta, \phi)}
[K(\theta, \phi)\sin\theta(n_1\cos\theta+\sin\theta(n_2\cos\phi+n_3\sin\phi))\\
\label{fdd3}
&&+\sin\theta(\cos\theta(n_2\cos\phi+n_3\sin\phi)-n_1\sin\theta)K,_\theta]+(n_3\cos\phi-n_2\sin\phi)K,_\phi
\Big\}\\
\nonumber
&&+\frac{\sqrt{2}q_0}{\kappa r}\csc^2(\theta/2)[{\cal L}(\theta, \phi)-\sin\theta {\cal L},_\theta].
\end{eqnarray}
where we have used (\ref{ep}). In this approximation is then easy to show that the nonvanishing components of ${\cal F}^{\mu\nu}$ -- also in second order of $1/r$ -- are given by
\begin{eqnarray}
\label{dual}
{\cal F}^{tr}\simeq F_{\theta\phi},~~{\cal F}^{t\theta}\simeq -F_{r\phi},~~{\cal F}^{\theta\phi}\simeq F_{tr}.    
\end{eqnarray}

Applying (\ref{fdd1})--(\ref{fdd3}) in (\ref{emf}) one may notice that
the electric field is purely radial, namely
\begin{eqnarray}
\label{el1}
E^\mu \simeq -\Big(\frac{\sqrt{2}q_0}{\kappa r^2}\Big) \delta^\mu_{~r}.   
\end{eqnarray}
It is worth noting that (\ref{el1}) is irrespective of the chosen boost direction so that
the axial symmetric electric field obtained in \cite{Aranha:2024aye} is the same in our general case. However, it is important to remark that in \cite{Aranha:2024aye} we have fixed a proper nonrotating frame of reference
which differs from our timelike observer fixed by (\ref{obs}). Apart from the fact such distinction does not alter the electric field, in the following we shall see that the magnetic counterpart exhibits a similar structure for both observers.

The substitution of (\ref{dual}) in (\ref{emf2}) is a rather involved task
in the sense that magnetic field generated deeply depends on the boost direction. Nonetheless, numerical simulations qualitatively show that
the magnitude of the resulting magnetic fields develops two intense lobes in the direction opposite to the boost. To illustrate this behaviour we choose four different boost directions in Fig. 2 fixing the same parameters as in reference \cite{Aranha:2024aye}. In these plots the color scales characterize the intensity of the modulus of the respective magnetic fields.
In the top panels we observe that the magnitude of the intensity appears collimated in the opposite direction of the boost, in the same manner as in  \cite{Aranha:2024aye}. This pattern is qualitatively maintained as we choose different boost directions as shown the in the bottom plots.  

\section{Final Remarks}

In this paper we have obtained an approximate solution of Einstein field equations which describes a boosted Kerr-Newman black hole relative to a Lorentz frame at future null infinity.
The boosted black hole geometry arises from 
a general twisting metric whose boost is given by the BMS
group. Employing a standard procedure we build the electromagnetic energy-momentum
tensor with the Kerr boosted metric together with its timelike Killing vector as the electromagnetic potential. We then show that Einstein field equations are satisfied -- up to fourth order in $1/r$ --
and the resulting spacetime is described by a general Kerr-Newman black hole whose boost points in a arbitrary direction. Using a Bondi-Sachs frame of reference we examine 
the formation of event horizon and ergosphere of the general boosted black hole. Considering a proper timelike observer we show that the electric field generated by the boosted black hole displays a pure radial
behaviour irrespective of the boost direction. The magnetic counterpart on the other hand develops an involved structure with two intense lobes of the
magnetic field observed in the direction opposite to the boost.

As a future perspective we intend apply the analysis presented in this paper to
different physical systems. The first thing to be addressed concerns configurations in which the electromagnetic fields generated by generally boosted black holes interact with surrounding matter, leading to the formation of a plasma magnetosphere. In this framework, analyzing the motion of accelerated test particles within the background defined by equation (\ref{gbbh}) is of significant physical interest, particularly in relation to Blandford-Znajek processes\cite{Blandford:1977ds}. Additionally, numerical findings from plasma simulations of black hole jets\cite{Parfrey:2018dnc} warrant examination within the broader context of generally boosted black holes.

Another application which deserves a careful examination is 
the formation of photon surfaces and strong gravitational lensing phenomena due to general boosted black holes. 
In fact, considering the boost effect we expect to obtain deformed photon surfaces
-- analogous to those of Kerr-Newman photon spheres -- in the boost direction. Given the breaking of axial symmetry of the background spacetime due to the inclusion of a lens described by an arbitrary boosted rotating black hole we also expect a shift of Einstein rings relative to the optical axis.

\section{Acknowledgments}

RM acknowledges financial support from
FAPERJ Grant No. E-$26/010.002481/2019$.

\end{document}